\documentclass[%
 reprint,
superscriptaddress,
nofootinbib,
 amsmath,amssymb,
 aps,
]{revtex4-2}

\usepackage[utf8]{inputenc}
\usepackage{siunitx}
\usepackage{hyperref}
\hypersetup{colorlinks,
citecolor=blue,
linkcolor=purple,
filecolor =purple,
anchorcolor = blue}
\usepackage{array}
\usepackage{physics}
\usepackage{dsfont}
\usepackage[english]{babel}			
\usepackage{xcolor}
\usepackage{graphicx}				
\usepackage{amssymb,amsmath}		
\usepackage{marvosym}				
\usepackage[T1]{fontenc}            %
\usepackage{wrapfig}				
\usepackage{charter} 				
\usepackage{blindtext}				
\usepackage{datetime}				
\usepackage{lipsum}                 
\usepackage{float}                  
\usepackage{tabularx}
\usepackage{tabulary}
\usepackage{dcolumn}
\usepackage{bm}

\newcommand{\UIBK}{Institute f{\"u}r Experimentalphysik, Universit{\"a}t Innsbruck, Innsbruck, Austria}
\newcommand{\WWU}{Institut f{\"u}r Festk{\"o}rpertheorie, Universit{\"a}t M{\"u}nster, 48149 M{\"u}nster, Germany}
\newcommand{\Bayreuth}{Theoretische Physik III, Universit{\"a}t Bayreuth, 95440 Bayreuth, Germany}
\newcommand{\JKU}{Institute of Semiconductor and Solid State Physics, Johannes Kepler University Linz, Linz, Austria}

\preprint{APS/123-QED}

\begin{document}

\title{SUPER Scheme in Action:\\Experimental Demonstration of Red-detuned Excitation of a Quantum Dot}


\author{Yusuf Karli }
    \thanks{These two authors contributed equally}
    \affiliation{\UIBK}
\author{Florian Kappe}
    \thanks{These two authors contributed equally}
    \affiliation{\UIBK}
\author{Vikas Remesh}%
\affiliation{\UIBK}
\author{Thomas K. Bracht}%
\affiliation{\WWU}
\author{Julian Muenzberg}%
\affiliation{\UIBK}
\author{Saimon Covre da Silva}%
\affiliation{\JKU}
\author{Tim Seidelmann}%
\affiliation{\Bayreuth}
\author{Vollrath Martin Axt}%
\affiliation{\Bayreuth}
\author{Armando Rastelli}%
\affiliation{\JKU}
\author{Doris E. Reiter}%
\affiliation{\WWU}
\author{Gregor Weihs}%
\affiliation{\UIBK}

\date{\today}

\begin{abstract}
The quest for the perfect single-photon source includes finding the optimal protocol for exciting the quantum emitter. Based on a recently proposed, so-called SUPER (swing-up of quantum emitter population) scheme, we demonstrate experimentally that two red-detuned laser pulses, neither of which could yield a significant upper-level population individually, lead to the coherent excitation of a semiconductor quantum dot. We characterize the emitted single photons and show that they have properties comparable to those achieved under resonant excitation schemes. 
\end{abstract}

\maketitle

\section{Introduction}
The future of photonic quantum technologies relies on bright, photostable, and on-demand sources of single and indistinguishable photons. In the search for such perfect quantum light sources, semiconductor quantum dots have emerged as a promising platform with excellent performance characteristics \cite{gazzano2013bright,Ding2016,senellart2017high,Schweickert2018,rodt2020deterministically,srocka2020deterministically,Wang2020,Tomm2021,Thomas2021b,lu2021quantum}. Quantum dots benefit from their excellent photostability, near Fourier-limited emission linewidth and growth technologies that allow easy integration into nanoscale devices offering scalability \cite{lodahl2015interfacing,dusanowski2020purcell,uppu2020scalable,elshaari2020hybrid,bounouar2020quantum}. 

To operate as an on-demand single-photon source, the quantum dot has to be prepared in its exciton state, i.e., a single electron-hole pair is generated in the quantum dot, which recombines to emit a single photon. One group of preparation schemes rely on above-band optical or electrical excitation, where carriers are generated in the barrier or the wetting layer that eventually relax to the quantum dot exciton state followed by single-photon emission \cite{buckley2012engineered}. These schemes are very appealing from a practical point of view, but have severe drawbacks due to preceding non-radiative relaxation processes, intrinsic time jitter, and charge fluctuations by excess carriers, which affect the properties of the emitted photons \cite{schlehahn2016generating}. Another group of excitation schemes employs two-photon excitation to the biexciton state \cite{jayakumar2014,prilmuller2018hyperentanglement,aumann2021demonstration, scholl2020crux, Huber2015}, which relaxes to the exciton state via the emission of a first single photon and emits a second single photon when decaying further to the ground state. In addition, several resonant or near-resonant excitation schemes have been developed \cite{luker2019review}. Examples are Rabi rotations \cite{stievater2001rabi,kamada2001exciton,ramsay2010damping,He2013}, chirped excitations exploiting the adiabatic rapid passage effect \cite{simon2011robust,wu2011population,mathew2014subpicosecond,wei2014deterministic,debnath2013high,Kaldewey2017b}, dichromatic excitation \cite{peiris2014bichromatic,he2019coherently,koong2021coherent} or phonon-assisted excitation \cite{Ardelt2014,bounouar2015phononassisted,quilter2015phononassisted,barth2016fast,cosacchi2019emission,Thomas2021b, huber2017highly}. Each scheme has its own advantages and disadvantages. For example, Rabi and chirped excitation schemes operate at photon energies resonant with the emitted photons and thus require challenging polarization filtering techniques. Dichromatic and phonon-assisted methods on the other hand are off-resonant enough to minimize or even avoid polarization filtering. While Rabi and dichromatic excitation are rather sensitive to the excitation pulse parameters, chirped and phonon-assisted schemes are known to be robust against those. Apart from the phonon-assisted scheme, which includes another particle and is thus incoherent, the other schemes are coherent.

In this paper, we use two red-detuned pulses in the recently proposed \cite{bracht2021swing} swing-up of quantum emitter population (SUPER) scheme to excite a quantum dot. It is rather surprising that this is possible at all, because neither of the two pulses alone will yield a significant exciton occupation. The SUPER scheme uses the coherent coupling of two laser fields to a quantum emitter to swing-up the system to full inversion. The decisive advantage over the dichromatic scheme \cite{peiris2014bichromatic,he2019coherently,koong2021coherent} is that both pulses are red-detuned and therefore no higher-lying states of the quantum dot will be directly addressed.  Because it operates below the band gap but yet coherently, this new scheme has the prospect of delivering high-quality single photons with only moderate experimental difficulty.


\section{Methods}
\label{sec:exp}
\begin{figure*}[ht]
    \includegraphics[width = 1\linewidth]{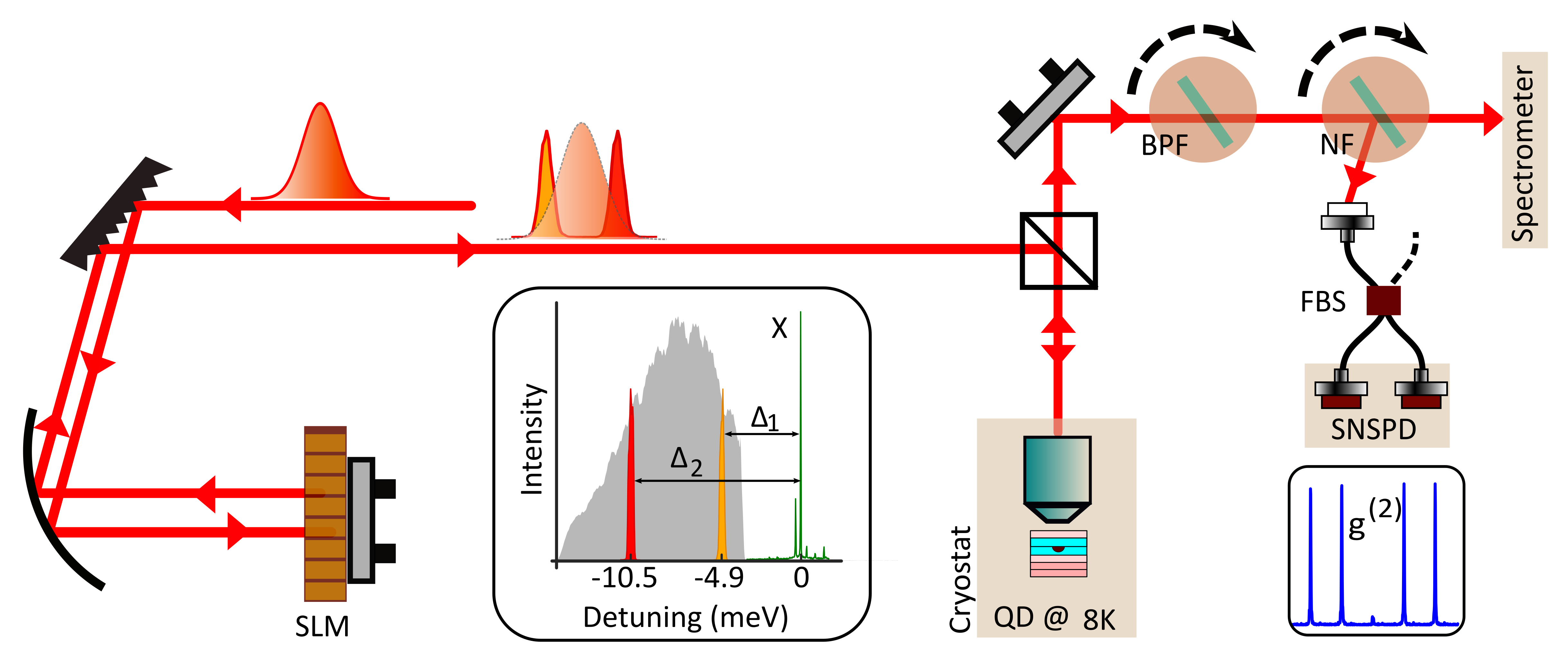}
    \caption{
    \textbf{Sketch of the experimental setup:} The laser beam is guided to a folded 4$f$ pulse shaper equipped with a spatial light modulator (SLM) for amplitude-shaping the broadband spectrum (grey shade, inset). Incoming and outgoing beams are shown as separate paths for clarity. The shaped pulse-pair is directed to the cryostat that holds the quantum dot at \SI{8}{\kelvin}. Emitted photons from the quantum dot are sent through a bandpass filter (BPF) and a notch filter (NF) to either the spectrometer or with an additional fiber beam splitter (FBS) to the superconducting nanowire single-photon detectors (SNSPD) to record the photon coincidences. Based on above-band excitation, the quantum dot exciton emission line (X, green) is identified. An exemplary pulse-pair with detunings $\Delta_1$ (orange) and $\Delta_2$ (red) are also shown.}

    \label{fig:Setup}
\end{figure*}

\subsection{SUPER scheme}
Here we briefly summarize the idea of the SUPER scheme \cite{bracht2021swing} for implementation using a semiconductor quantum dot. Assuming that the quantum dot can be approximated as a two-level system consisting of the ground and exciton state, the SUPER scheme requires two laser pulses, that are both red-detuned ($\Delta_1$ and $\Delta_2$) from the quantum dot exciton state by several millielectronvolts. Considering a Gaussian-shaped excitation pulse, we define the generalized Rabi frequency as $\hbar \Omega_{i}^{\text{Rabi}}=\sqrt{(\hbar \Omega_i)^2+\Delta_i^2}$, where $\Omega_i$ is the resonant Rabi frequency given at the maximum of the pulse envelope. In the SUPER scheme, one achieves a gradual rise in the exciton population by modulating the Rabi frequency, through the beating of the two detuned pulses. According to Ref.~\cite{bracht2021swing}, if the difference between the two detunings coincides with the Rabi frequency $|\Delta_2-\Delta_1| = \hbar \Omega_{1}^{\text{Rabi}}$, implying the condition $|\Delta_2| >  2|\Delta_1|$, the SUPER mechanism results in a complete population inversion.


\subsection{Pulse-pair generation}

To generate red-detuned pulses with appropriate detunings, we rely on amplitude-shaping with a programmable spatial light modulator (SLM, CRi Inc., 128 pixels, single mask). The experimental scheme is summarized in Figure~\ref{fig:Setup}. A broadband Ti:Sapphire laser (MIRA 900, Coherent Inc.) produces \SI{300}{\femto\second} long, Gaussian-shaped pulses with the central wavelength of \SI{802}{\nano\meter}, pulse energy of $\approx \SI{4}{\nano\joule}$ and a peak power of $\approx \SI{12}{\kilo\watt}$. With a home-built, folded 4$f$ pulse shaper equipped with the SLM, two phase-locked pulses are amplitude-shaped out of the initial spectrum. To this end, the collimated laser beam that enters the 4$f$ pulse shaper is first dispersed by a blazed diffraction grating (1200 lines/mm, Newport), and then focused onto the SLM with a curved mirror ($f= \SI{500}{\milli\meter}$), such that each pixel is assigned a narrow laser spectral window ($\approx  \SI{0.05}{\nano\meter}$). The laser beam then gets reflected back by the mirror behind the SLM through the same path and leaves the pulse shaper. For a detailed description, see Sec.~\ref{app:specifics} in the Appendix. We can calibrate the relationship between the drive level of the SLM pixels and the resultant attenuation function \cite{monmayrant2010newcomer} to create a pulse pair with various detunings and amplitudes. The resolution of our setup allows us to tune the frequency of the pulses on a sub-millielectronvolt scale. 
 
A representative amplitude-shaped spectrum to excite the quantum dot is shown in the inset of Figure~\ref{fig:Setup}. To characterize the detuning of the two pulses with respect to the exciton line, we first performed above-band-gap excitation of the quantum dot. The resulting emission spectrum is shown as a green curve in Figure~\ref{fig:Setup}. A sharp exciton-emission line (X) is identified at \SI{798.66}{\nano\meter}, surrounded by phonon sidebands and substrate emission. From this we can estimate the detunings $\Delta_1$ and $\Delta_2$. In the following, we will always refer to the pulse with the smaller detuning as first pulse i.e., $|\Delta_1|<|\Delta_2|$. Figure~\ref{fig:Setup} also shows the unshaped laser spectrum as grey-shaded. The sharp edge near the first pulse is due to a razor blade mounted behind the SLM to suppress the laser spectral tail that is resonant with the quantum dot emission line. The intensities of both pulses can be tuned individually by varying the transmissivities of the SLM pixels from 0 to 1, denoted as $T_1$ and $T_2$. For our experiments, the intensities of the second pulse at $\Delta_2 = \SI{-10.6}{\milli\electronvolt}$ range from \SI{0.7}{\micro\watt} ($T_2 = 0$) to \SI{18.8}{\micro\watt} ($T_2 = 1$) while the first pulse intensity was fixed to \SI{15.5}{\micro\watt}, measured at the cryostat window.

\begin{figure*}[!htbp]
    \centering
    \includegraphics[width =\linewidth]{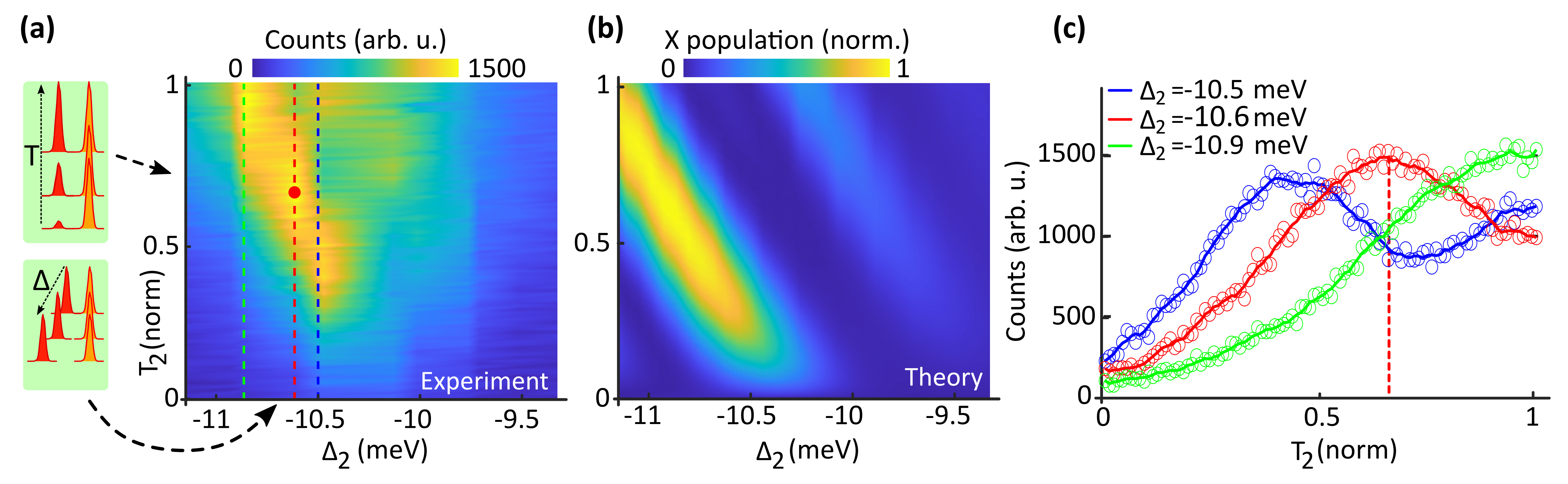}
    \caption{\textbf{Exciton probability achieved by the SUPER scheme:} (a) Measured photon counts at exciton emission energy as a function of the detuning $\Delta_2$ and the transmissivity $T_2$ of the second pulse. The first pulse is fixed to $\Delta_1= \SI{-4.9}{\milli\electronvolt}$ and $T_1=0.5$. The scale shows the integrated exciton counts after correcting for background. The red dot indicates the parameters used in the photon quality experiments (Section \ref{sec:Photon_Q} in the Appendix). (b) Theoretically calculated exciton (X) population based on a two-level system. (c) Cuts through the 2D map for $\Delta_2=\SI{-10.5}{\milli\electronvolt}$ (blue), $\SI{-10.6}{\milli\electronvolt}$ (red) and $\SI{-10.9}{\milli\electronvolt}$ (green) as indicated by vertical lines in (a). The vertical red line indicates the parameters at which the photon quality measurements are performed. 
    }
    \label{fig:HM1}
\end{figure*}
\subsection{Quantum dot measurements}
\label{subsec:QDmeasurements}

The shaped laser pulse is then directed to the closed-cycle cryostat (base temperature \SI{8}{\kelvin}), where the quantum dot sample is mounted on a three-axis piezoelectric stage (ANPx101/ANPz102, Attocube Systems AG). Our sample consists GaAs/AlGaAs quantum dots obtained by the Al-droplet etching method \cite{huber2017highly,da2021gaas}. The dots are embedded in the center of a lambda-cavity placed between a bottom (top) distributed Bragg reflector consisting of 9 (2) pairs of  $\lambda/4$ thick Al$_{0.95}$Ga$_{0.05}$As/Al$_{0.2}$Ga$_{0.8}$As layers. The quantum dot emission wavelength is centered around 800 nm.The excitation laser is focused onto the quantum dots with a cold aspheric lens (NA = 0.77, Edmund Optics). 

The quantum dot emission is collected via the same path backwards, where a combination of a bandpass filter (\SI{808}{\nano\meter}, FWHM \SI{3}{\nano\meter}, Layertec) and a notch filter (BNF-805-OD3, FWHM \SI{0.3}{\nano\meter}, Optigrate) help separate the exciton emission from the background and residual scattered laser light. The collected photons are routed to a single-photon sensitive spectrometer (Acton SP-2750, Roper Scientific) equipped with a liquid Nitrogen cooled charge-coupled device camera (Spec10 CCD, Princeton Instruments) or superconducting nanowire single-photon detectors (SNSPD, Single Quantum). We record the emitted spectra for \SI{250}{\milli\second} integration time. For estimating the wavelength independent background, we integrate the photon counts on the high-energy sideband of the exciton peak (for a detailed discussion, see Sec.~\ref{app:background} in the Appendix).

\section{Results}


\subsection{Exciton population under SUPER excitation} 
To start, we fix the detuning and the transmissivity of the first pulse to $\Delta_1=\SI{-4.9}{\milli\electronvolt}$ and $T_1=0.5$, respectively. We then vary the transmissivity of the second pulse ($T_2$), and record the emitted spectra for various detunings ($\Delta_2$). The results are displayed as a two-dimensional map in Figure~\ref{fig:HM1}(a), as a function of $\Delta_2$ and $T_2$, where the color scale denotes the integrated photon counts. All data shown is corrected in respect to a background signal obtained via the method described in Sec. \ref{app:background} in the Appendix.   
Every automated intensity scan (i.e., individual columns in Figure~\ref{fig:HM1}(a)) records emitted spectra for 100 $T_2$ values, and the experiment is performed for eleven different $\Delta_2$ values. At zero intensity of the second pulse, i.e., $T_2=0$, only negligible photon counts are recorded, implying that no excitation occurs in the absence of the second pulse, even if the first pulse is present. With increasing $T_2$ from 0 to 1, the exciton counts gradually increase, specifically for detunings around $\Delta_2=-10\ldots \SI{-11}{\milli\electronvolt}$. We find a clear region of high photon counts demonstrating that the exciton state gets occupied by the two-pulse excitation. To validate further that the exciton state only gets populated when both pulses are switched on, we set $T_1=0$, and perform the $T_2 - \Delta_2$ scan, as in Figure~\ref{fig:HM1}(a)), which does not result in any significant exciton emission (see Figure~\ref{fig:single_pulse} (a) in the Appendix). 

Therefore, we conclude that under the action of both pulses, an excitation via the SUPER, below bandgap, excitation has taken place. To gain further insight into this, we calculate the dynamics of the two-level system under the double pulse excitation, and the results are shown in Figure~\ref{fig:HM1}(b) (for details of the modeling see Sec.~\ref{app:theory} in the Appendix. The observed high exciton occupation at $\Delta_2\approx \SI{-10.5}{\milli\electronvolt}$ with a diagonal trend towards larger $\Delta_2$ and $T_2$, agree with the theoretical results. The calculated maximum exciton occupation is $\approx$ 97~\% in the considered parameter window. Interestingly, for both experiment and theory, the condition that $|\Delta_2|>2|\Delta_1|$ holds. 

Figure~\ref{fig:HM1} (c) shows line plots of the measured exciton occupation for $\Delta_2= \SI{-10.5}{\milli\electronvolt}$, $\SI{-10.6}{\milli\electronvolt}$ and $\SI{-10.9}{\milli\electronvolt}$. 
From Figure~\ref{fig:HM1} (c) we can see another interesting behaviour: For the largest detuning $\Delta_2=\SI{-10.9}{\milli\electronvolt}$ (green line) we find that the exciton counts increase monotonically with increasing $T_2$. For $\Delta_2=\SI{-10.6}{\milli\electronvolt}$ (red line), we find an increase in exciton counts up to $T_2=0.64$, after which it decreases again. The most striking observation is for $\Delta_2=\SI{-10.5}{\milli\electronvolt}$ (blue line), which shows up to 1.5 oscillations from $T_1=0\ldots1$ with a maximum at $T_2 = 0.4$ and a minimum at $T_2 = 0.7$. All these findings provide compelling evidence that the recorded exciton emission is due to the coherent excitation with two red-detuned pulses.

We observe similar results for a few different fixed-intensities of the first pulse $T_1$ (see Figure~\ref{fig:T1_scan} in the Appendix). We do not record any significant exciton counts for $T_1 < 0.4$ for the $T_2 - \Delta_2$ two-dimensional scans as described before. Thus, our results reveal that the optimum parameter range for the coherent control of exciton population via the SUPER scheme in our experiments is around $\Delta_2=\SI{-10}{\milli\electronvolt}$ and $T_1$ around 0.5 (see Sec.~\ref{app:specifics} in the Appendix for more details). 

\subsection{Photon quality measurements}
\label{sec:Photon_Q}

\begin{figure}[ht]
    \centering
    \includegraphics[width = 1\linewidth]{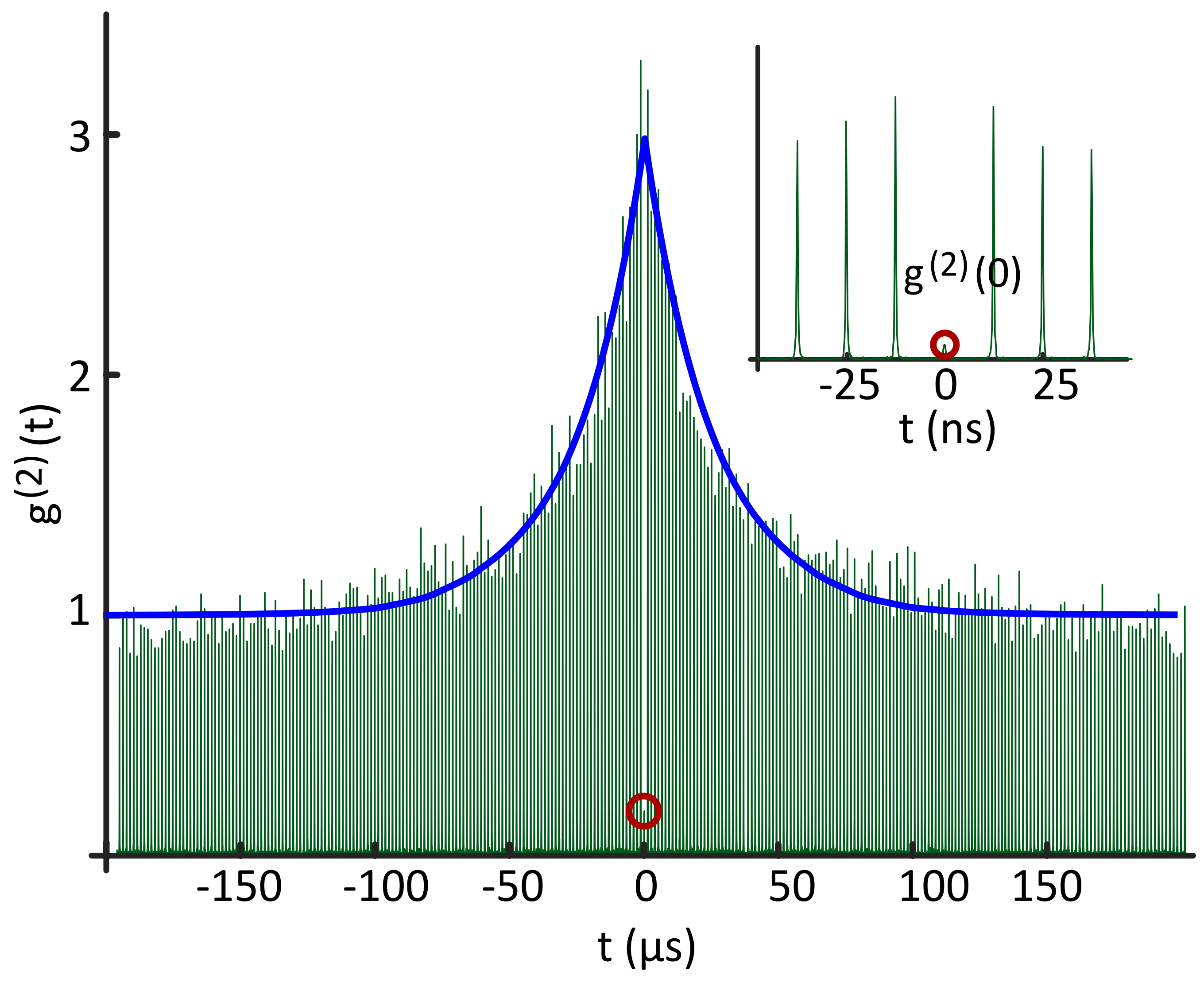}
    \caption{\textbf{Second order correlation function measurement ($\bm{g^{(2)}(t)}$).} Obtained $g^{(2)}(t)$ (green curves) by performing an HBT-measurement and fitting a converging exponential function (blue solid curve). The red circle indicates the peak height at $t=0$. A background-corrected $g^{(2)}_{\text{corr}}(0) = 0.02(6)$ is obtained. The inset shows the data around $t=0$.} 
    \label{fig:g2}
\end{figure}

Based on the aforementioned results, we choose $\Delta_1=\SI{-4.9}{\milli\electronvolt}$ and $T_1=0.5$ as well as $\Delta_2=\SI{-10.6}{\milli\electronvolt}$ and $T_2=0.64$, where the largest exciton emission is recorded (indicated by the red dot in Figure~\ref{fig:HM1}), as the excitation parameters to investigate the photon quality. Firstly, we validate the nature of the emission by measuring the decay dynamics. For this, we send the emitted photons to the SNSPD for the time-correlated single-photon counting measurement with \SI{20}{\pico\second} temporal resolution. An internal photodiode within the excitation laser serves as the start signal, while single photon clicks act as stop signals. The resulting time-correlated histogram is fit with an exponential function, which leads to a computed lifetime of \SI{165.1(6)}{\pico\second} (see Sec. \ref{app:LifeTime} in the Appendix). This is in excellent agreement with our results under resonant excitation of the same quantum dot \cite{Munzberg2022}. 

We then verify the single-photon characteristics by measuring the second-order correlation $g^{(2)}{(t)}$ in a Hanbury Brown and Twiss (HBT) setup. For this, we send the emitted photons to two SNSPD channels located at the two outputs of a 50/50 fiber beam splitter (see Figure~\ref{fig:Setup}) and record their temporal correlations with a time bin width of \SI{100}{\pico\second}.
The corresponding results are shown in Figure~\ref{fig:g2}. We clearly see that the computed $g^{(2)}(0)$ is well below $0.5$, which proves the single-photon character. 

In the literature, various methods exist to quantify $g^{(2)}(0)$. Firstly, it is usual that $g^{(2)}(t)$ for light emitted from a quantum dot drops down exponentially to 1 (at $|t|>>0$), which is attributed to blinking (on-off jumps) of the quantum dot \cite{vuvckovic2003enhanced,santori2004submicrosecond, Davanco}. Therefore, we apply an exponential fitting function $a+b\cdot{\exp}(-|t|/c)$ to the recorded coincidence counts and normalize to $a$ (blue line in Figure~\ref{fig:g2}). From this, we compute $g^{(2)}_{\text{meas}}(0) = 0.19(3)$.
Next, we estimate the background contribution in the recorded exciton photon counts (see Sec.~\ref{app:background} in the Appendix for detailed information). 
We define $g^{(2)}_{\text{corr}}(0) = 1+ (g^{(2)}_{\text{meas}}(0) -1)/\rho^2$ where $\rho$ is the ratio of the signal to the total counts \cite{brouri2000photon,peter2007fast}. Based on the estimated background contribution (see Sec.~\ref{app:background} in the Appendix), we calculate $\rho = 0.91(1)$, resulting in $g^{(2)}_{\text{corr}}(0) = 0.02(6)$. Lastly, if we take the ratio of the integrated area of the Gaussian fit of the coincidence peak at \SI{-25}{\nano\second} to that at \SI{0}{\nano\second}, we find that $g_{\text{post}}^{(2)}(0)=0.073(3)$. This value is meaningful if one post-selects the emitted photons from the quantum dot based on its on-off timescales. Notably, on resonant excitation of the same quantum dot, we obtained $g^{(2)}_{\text{res}}(0) = 0.016(1)$ \cite{Munzberg2022}. Thus, we establish that the quality of single photons produced in the SUPER scheme is on par with that of the coherent excitation scheme. We summarize the obtained values of $g^{(2)}(0)$ in Table~\ref{tab:g2}.
\begin{table}[H]
    \centering
    \caption{Summary of the measured and computed $g^{(2)}(0)$ values.}
    \begin{tabular}{lcS}
        \text{\textbf{Method}} & \text{\textbf{Symbol}} & \text{\textbf{Value}} \\
        \hline
        Raw data & $~~~~g^{(2)}_{\text{meas}}(0)~~~~$ & 0.19(3) \\
        Background corrected&$g^{(2)}_{\text{corr}}(0)$ & 0.02(6) \\
        Peak area ratio$~~~~$ & $g_{\text{post}}^{(2)}(0)$ & 0.073(3) \\
        Resonant excitation &$g^{(2)}_{\text{res}}(0)$ & 0.016(1) \\
        
    \end{tabular}
    \label{tab:g2}
\end{table}



\section{Conclusion}
To summarize, this work demonstrates that, under optimized detuning and intensity, a red-detuned, phase-locked pulse pair can populate the exciton state in a quantum dot relying on the SUPER mechanism. We also perform photon quality measurements and contrast the results with that of resonant excitation. Our results contribute towards an effortless method for generating high purity single photons, with the proof-of-concept performance on par with the resonant excitation scheme, yet most importantly, removing the need for stringent polarization filtering. The red detuned, pulse-pair excitation could have wider implications, especially for probing the coherent dynamics of multi-excitonic systems.  

\section*{Acknowledgements}
The authors acknowledge Marita Wagner for assistance and Ron Stepanek (Meadowlark Optics Inc.) for valuable discussions during the experiment. YK acknowledges the financial support from the Doctoral Program W1259 (DK-ALM Atoms, Light, and Molecules). VR acknowledges the financial support from the Austrian Science Fund FWF projects TAI-556N (DarkEneT) and I4380 (AEQuDot). TKB and DER acknowledge financial support from the German Research Foundation DFG through project 428026575 (AEQuDot). AR and SFCdS acknowledge C. Schimpf for fruitful discussions, the FWF projects FG 5, P 30459, I 4320,  the Linz Institute of Technology (LIT) and the European Union's Horizon 2020 research, and innovation program under Grant Agreement Nos. 899814 (Qurope), 871130 (ASCENT+).

\bibliography{ref_v3.bib}

\clearpage

\clearpage
\section{Supplementary information}

\subsection{Experimental details} \label{app:specifics}

In addition to the description of the set-up in Sec.~\ref{sec:exp}, here we provide more information on the experiment. 
The broadband laser used for the experiments has a pulse duration ($\tau$) of $\SI{300}{\femto\second}$ and a repetition frequency ($f_r$) of \SI{76}{\mega\hertz} at a central wavelength of \SI{802}{\nano\meter}. For Gaussian shaped pulses, the peak power ($P_{\text{peak}}$) is defined as the ratio of the pulse energy ($E_{\text{peak}}$) and $\tau$, as $P_{\text{peak}} = 0.94 E_{\text{peak}}/\tau$, where $E_{\text{peak}}$ is the ratio of average power ($P^{\text{avg}}$) and $f_r$. For the experiments described in this manuscript, the maximum $P^{\text{avg}}$ measured before the entrance of the 4$f$ pulse shaper is \SI{300}{\milli\watt}, that corresponds to $\SI{4}{\nano\joule}$ energy per pulse and a peak pulse power of $\SI{12}{\kilo\watt}$. 

The home-built 4$f$ pulse shaper constructed in folded (reflection) geometry includes a programmable, liquid crystal on silicon spatial light modulator (SLM-128-A-VN, CRi Inc., 128 pixels, single mask). Here, the incoming laser beam first hits a grating (1200 lines/mm) and disperses, before being focused onto the SLM that is kept at the Fourier plane of a curved mirror of $f = \SI{500}{\milli\meter}$. Effectively, each pixel of the SLM holds a narrow spectral region of the dispersed laser beam. A flat mirror kept behind the SLM that is tilted slightly on the vertical axis, provides vertical displacement to pick up the reflected beam above the incoming path. The distance traveled by the laser beam within the shaper amounts strictly to 4$f$, which ensures that the laser pulses leave the pulse shaper free of dispersion \cite{weiner2000femtosecond}. We verify this condition through a nonlinear measurement of the pulse duration by an autocorrelator (PulseCheck, APE GmBH). Next, we calibrate the relationship between the applied voltage drive level and the resultant phase retardance of the liquid crystals based on the standard procedure \cite{monmayrant2010newcomer}. The achievable amplitude modulation is proportional to the cosine of the retardance of the liquid crystal cells. This enables the generation of a phase-locked pair of pump pulses with arbitrary amplitudes and detunings $\Delta_1$ and $\Delta_2$ as described in the main text. Our setup has a resolution of \SI{0.05}{\nano\meter} in wavelength (\SI{0.1}{\milli\electronvolt} in energy). 

We also note that broadband \si{\femto\second} pulses experience group delay dispersion while passing through high index media, e.g. single mode fibers, that results in their temporal stretching, and requires a spectral phase dispersion compensation before the focusing lens or at the focus \cite{accanto2014phase,gamouras2013simultaneous}. However, the calculated temporal duration of the detuned pulse pair in this experiment is $\approx \SI{7}{\pico\second}$, and the combined phase distortion due to the optical fiber that delivers the laser beam to the cryostat and the focusing aspheric lens is $\approx \SI{5000}{\femto\second\squared}$. Therefore, the resultant temporal stretching is negligible.

The quantum dots chosen for this experiment are grown by the droplet epitaxy technique \cite{huber2017highly,da2021gaas}. This process results in dots grown in random locations on the sample. To locate a bright quantum dot, we illuminate the sample with an above band laser beam (\SI{532}{\nano\meter}, Thorlabs) and optimize the position with the three-axis piezo stack. The focal spot size and the density of the dots in the sample ensure that no two dots are excited in a single excitation spot.

The integration time while acquiring the exciton emission is carefully chosen to avoid conflicts with the SLM frame transfer rate. Furthermore, a time delay of \SI{200}{\milli\second}, together with a feedback response following the spectral acquisition, ensures that measured spectra are free of unexpected artifacts. All the experiments are fully automated through a LabVIEW program.

\subsection{Theoretical model and calculations} \label{app:theory}
For the theoretical calculations, the quantum dot is approximated by a two-level system consisting of ground state $\ket{g}$ and excited state $\ket{x}$, which are separated by an energy $\hbar\omega_0$. This system is driven by a pulsed laser, given by a time-dependent term $\Omega(t)$ described in dipole and rotating wave approximation. The Hamiltonian for this system is
\begin{equation}
H = \hbar\omega_0\ket{x}\bra{x} - \frac{\hbar}{2}\left(\Omega^*\ket{g}\bra{x} + \Omega\ket{x}\bra{g}\right).
\end{equation}
Now, instead of specifying the single laser pulse parameters, the laser $\Omega(t)$ is obtained by inverse Fourier transform of the laser spectrum, multiplied by an amplitude mask to mimic the SLM. The broad spectrum before the SLM is approximated by a Gaussian frequency spectrum as,
\begin{equation}
    \tilde{\Omega}(\omega) = \frac{\tilde{\alpha}}{\sqrt{2\pi}\tilde{\sigma}}\exp(-\frac{(\omega-\omega_{\mathrm{C}})^2}{2\tilde{\sigma}^2})
\end{equation} with a spectral FWHM of $2\sqrt{2\ln 2}\hbar\tilde{\sigma} = \SI{17}{meV}$, centered around a detuning of $\hbar(\omega_{{C}}-\omega_0) = \SI{-8.4}{meV}$ and an integrated pulse area of $\tilde{\alpha} = 470\pi$. The SLM amplitude function is approximated by two normalized Gaussian shape functions,
\begin{equation}
    f_{\text{mask}}(\omega) = \tilde{T}_1 \exp(-\frac{(\omega-\omega_1)^2}{2\tilde{\sigma}_1^2}) + \tilde{T}_2 \exp(-\frac{(\omega-\omega_2)^2}{2\tilde{\sigma}_2^2}).
\end{equation}

\begin{figure}[hbt]
    \centering
    \includegraphics[width = 1\linewidth]{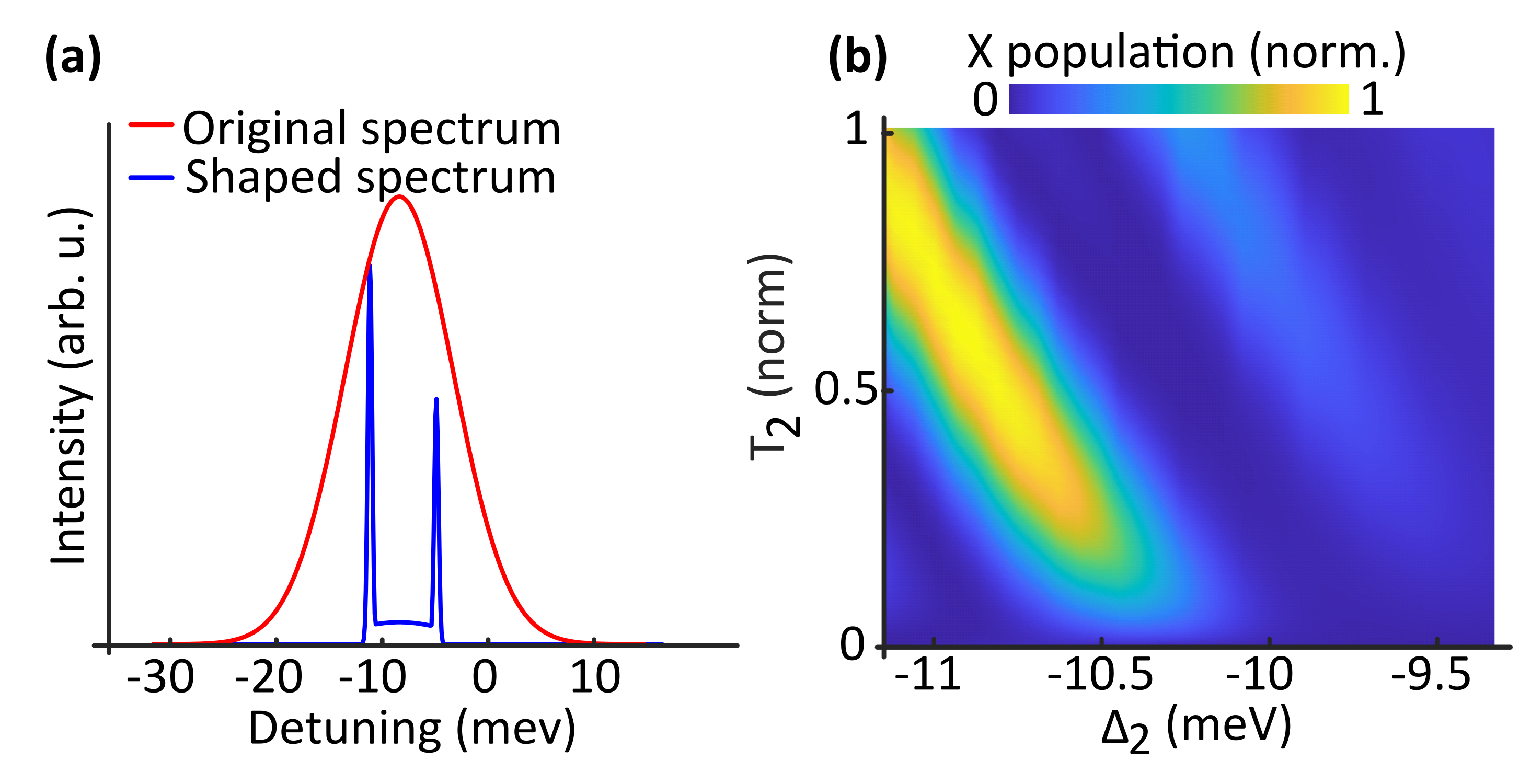}
    \caption{\textbf{Theoretical simulations:} (a) Spectrum of the broadband pulse (red) and an exemplary amplitude-shaped spectrum after the SLM (blue). (b) Calculated exciton population on SUPER excitation, considering a three-level system including exciton and biexciton states.}
    \label{fig:theory}
\end{figure}

This results in two contributions to the resulting spectrum at $\omega_i$, with variable transmission $\tilde{T}_{i}$. From this we define the detunings as $\Delta_i = \hbar(\omega_i - \omega_0)$. The spectral width $\hbar\tilde{\sigma}_i$ is chosen to \SI{0.2}{meV} for all calculations. Additionally, the mask has a transmission of \SI{5}{\percent} between the two peaks, accounting for the non-zero extinction of SLM pixels. A representative amplitude-shaped spectrum is shown in Figure~\ref{fig:theory}(a). Note that the transmission $\tilde{T}_i$ is for the electric field, in contrast to the experimental scenario where $T_i$ acts on the intensity, i.e., $T_2$ (the y-axis on Figure~\ref{fig:HM1}) corresponds to $(\tilde{T}_2)^2$. \\ To calculate the exciton occupation, we derive the equations of motion from the Hamiltonian using the von-Neumann equation, which is then numerically integrated \cite{bracht2021swing}.

Albeit a simplified theoretical model of the two-level system, we observe an excellent agreement between the theory and the experiment (cf.~Figure~\ref{fig:HM1}). Since the phonon influence on the SUPER scheme has been shown to be rather negligible \cite{bracht2021swing,bracht2022phonon}, we ignored it in our model. Since the excitation takes place with linear polarized pulses, we have additionally performed calculations within the standard three-level system including an additional biexciton state \cite{Kaldewey2017b}, with a biexciton binding energy of $E_B=\SI{4}{\milli\electronvolt}$. We observe that the SUPER scheme can be successfully applied in this system too (Figure~\ref{fig:theory}(b)), where the exciton population is calculated for $T_2 - \Delta_2$ scan, as in Figure~\ref{fig:HM1} (b). Here, $\tilde{\alpha}$ is set to $600\pi$ while all the other parameters remain the same as for the two-level system. The increased pulse area is necessary to compensate for the Stark shifts induced by the additional level, to still meet the condition for the SUPER scheme.

\subsection{Background estimation} 
\label{app:background}

\begin{figure}[hbt]
    \centering
    \includegraphics[width = 1\linewidth]{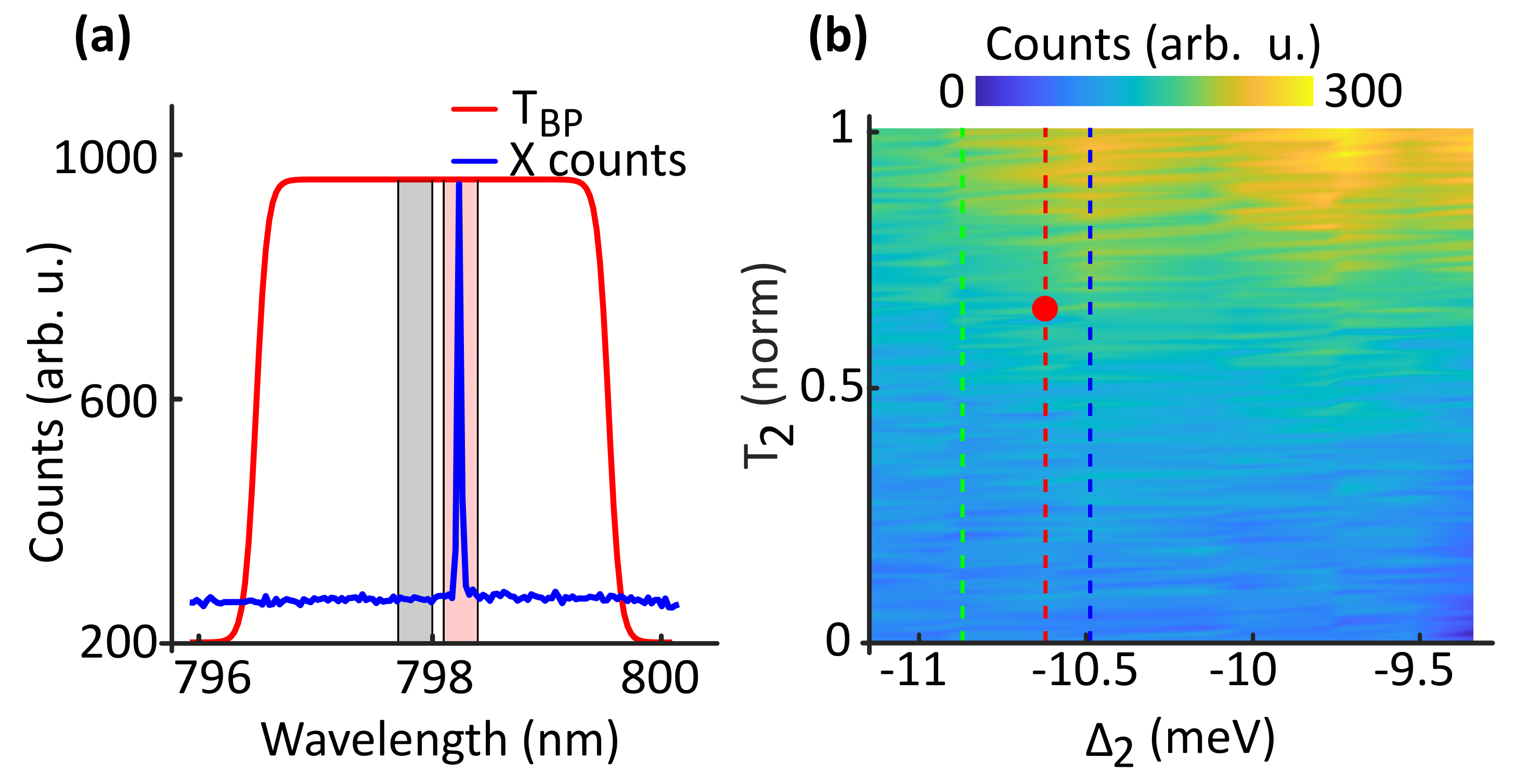}
    \caption{\textbf{Data processing and background estimation:} (a) A representative exciton emission spectrum (blue solid line) measured in the experiment; transmission spectrum of the bandpass filter used (solid red line) . The shaded red (grey) area is used for extracting the exciton (background) photon counts. (b) Background contribution in the $T_2 - \Delta_2$ scan, cf. Figure~\ref{fig:HM1}. Red dot denotes the  parameters for photon quality measurements, cf. Sec.~\ref{sec:Photon_Q}, yielding a \SI{9(1)}{\percent} background contribution. }
    \label{fig:background}
\end{figure}

In the experiments, the emitted light from the quantum dot is sent to the spectrometer, after passing through a bandpass filter with FWHM of \SI{3}{\nano\meter} at a central wavelength of \SI{798}{\nano\meter} (red solid curve in Figure~\ref{fig:background} (a)). An exemplary emission spectrum shown as blue solid curve in Figure~\ref{fig:background} (a) suggests the presence of a wavelength-independent background around the exciton emission line, which we attribute to fluorescence of the host material under high power illumination. For the results displayed in Figure~\ref{fig:HM1}, the background contribution is estimated from the integrated photon counts in the grey-shaded window and is then subtracted from the integrated photon counts in the red-shaded window (Figure~\ref{fig:background} (a)). The shaded windows correspond to \SI{0.3}{\nano\meter} spectral width, determined by the FWHM of the notch filter through which the exciton photons are sent to the SNSPD. In Figure~\ref{fig:background} (b) we show the variation in the background corresponding to the results displayed in Figure~\ref{fig:HM1}. On comparison, we estimate a \SI{9(1)}{\percent} contribution of background noise in the photon quality measurements (red dot in Figure~\ref{fig:background}). 

\subsection{Normalization of the pulse intensities}
\label{app:int_corr}

\begin{figure}[bt]
    \centering
    \includegraphics[width = 1\linewidth]{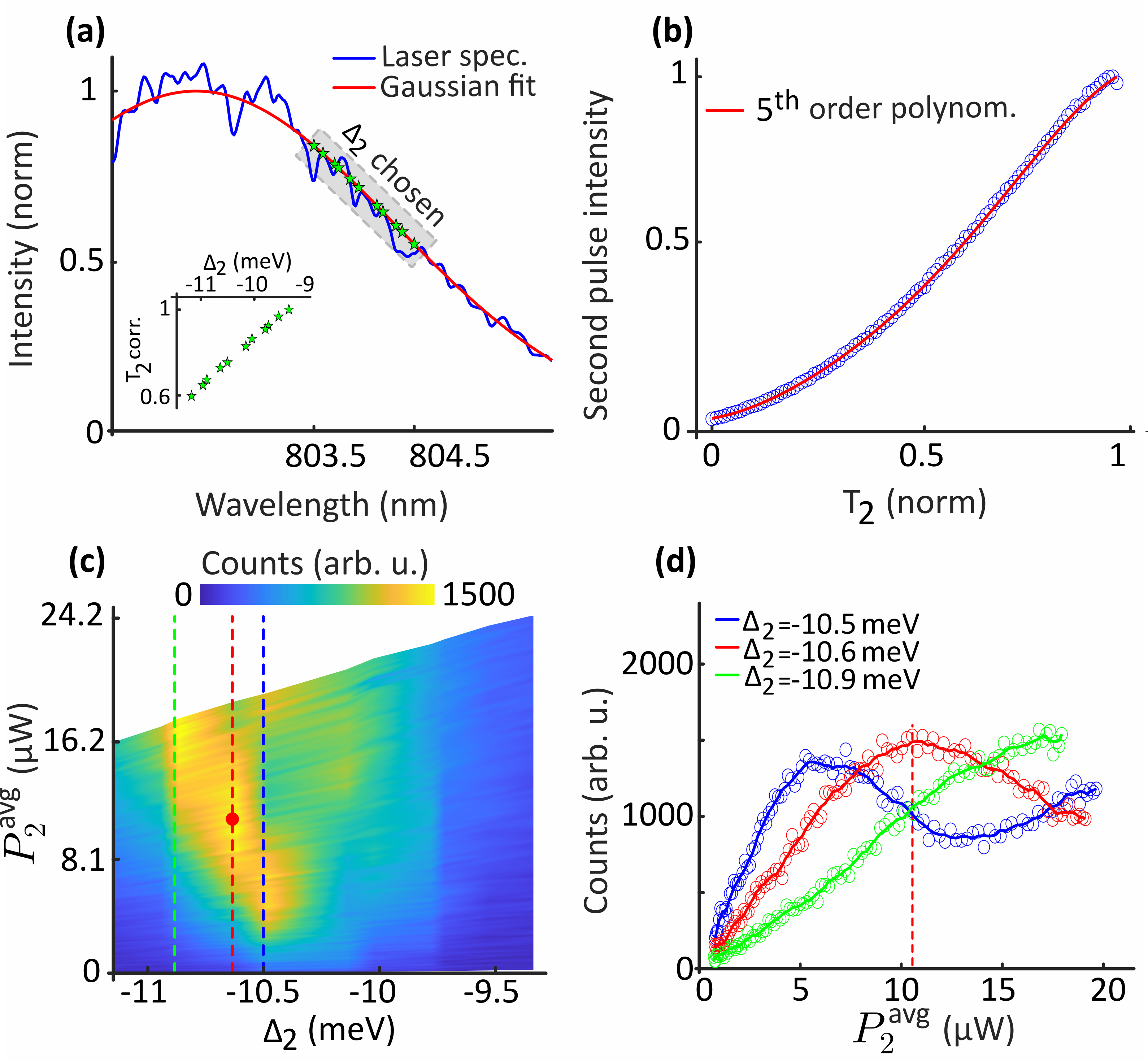}
    \caption{\textbf{Corrected transmissivities of individual pulses:} (a) Measured excitation laser spectrum (blue solid line), fit with a Gaussian function (red solid line) where the eleven chosen $\Delta_2$ values for the experiments are indicated (green stars, grey shaded and inset). (b) Transfer function to convert $T_2$ to the normalized intensity of the second pulse (blue circles). Red solid line is a fifth order polynomial fit. (c) and (d) Corrected Figure~\ref{fig:HM1}(a) and \ref{fig:HM1}(c) for $T_2$ values calibrated for $P^\text{avg}_2$ of the second pulse. The first pulse intensity ($T_1 = 0.5$ ) is \SI{15.5}{\micro\watt}.}
    \label{fig:T2_calib}
\end{figure}

For the experiment, the pulse pairs (with various $\Delta_2$ and $T_2$) are amplitude-shaped from the broadband laser spectrum as explained in the main text. This means that the individual intensities of the detuned pair 
depend on their spectral locations in the unshaped Gaussian intensity spectrum. In other words, the maximum of $T_2$ 
will not be the same for different $\Delta_2$. We therefore apply a correction procedure to normalize $T_2$. We apply Gaussian fits to the measured spectra for various detunings $\Delta_2$, compute the integrated spectral intensities, and obtain ${T_2\text{corr}.}$ with respect to a maximum transmission of ${T_2} = 1$ for $\Delta_2 = \SI{-9.4}{\milli\electronvolt}$ (see Figure~\ref{fig:T2_calib}, inset). 

We then normalize $T_2$ with relative pulse intensities based on a fifth order polynomial fit, as shown in Figure~\ref{fig:T2_calib}. Based on this fit function, we calibrate transmissivities $T_1$($T_2$) with respect to the measured power values $P^\text{avg}_1$($P^\text{avg}_2$), as summarized in the table \ref{tab:T_vs_Power}. The results presented in Figure~\ref{fig:HM1}(a) and \ref{fig:HM1}(c) will, as a result, get modified to \ref{fig:T2_calib}(c) and (d), where the $T_2$ axis is replaced by $P^\text{avg}_2$ values. 

\begin{table}[H]
    \centering
    \caption{Average power of the pulses $\Delta_1 = \SI{-4.9}{\milli\electronvolt}$ and $\Delta_2 = \SI{-10.6}{\milli\electronvolt}$, measured at the cryostat entrance window.}
    \begin{tabular}{SS|SS}
        \text{$\bm{T_1}$} & \text{$\bm{P^\text{avg}_1}$ \textbf{(\si{\micro\watt})}} & \text{$\bm{T_2}$} & \text{$\bm{P^\text{avg}_2}$ \textbf{(\si{\micro\watt})}} \\
        \hline
        0 & 0.4 & 0 & 0.7\\
        0.5 & 15.5  $~~~~~~~~~$& 0.65 & 10.9\\
        1 & 34.2 & 1 & 18.6\\
        
    \end{tabular}
    \label{tab:T_vs_Power}
\end{table}

   
    %
 
\subsection{Verification of the two-pulse effect}
\label{app:verification}

\begin{figure}[hbt]
    \centering
    \includegraphics[width = 1\linewidth]{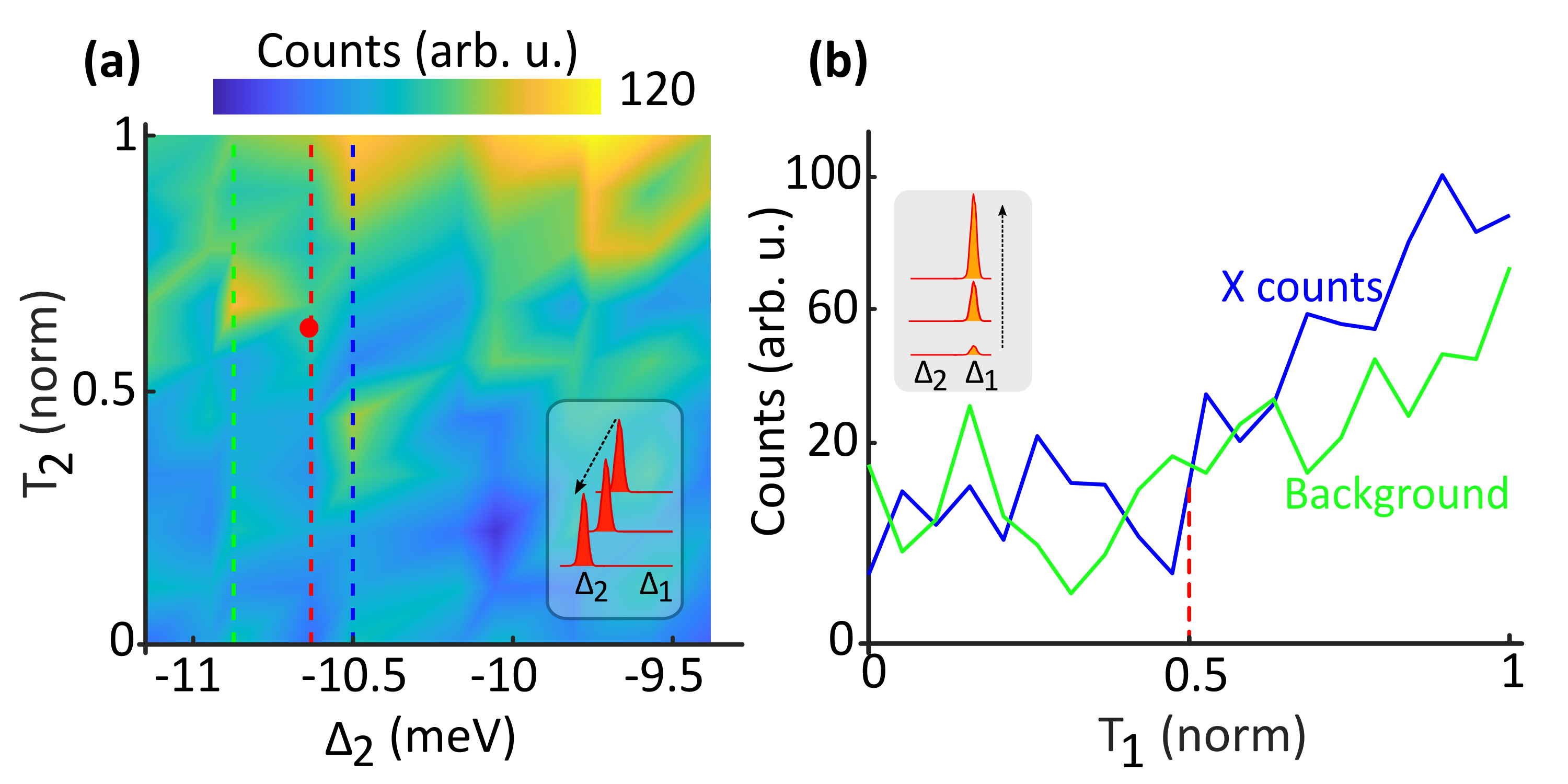}
    \caption{\textbf{Single pulse excitation:} (a) Measured photon counts at exciton-emission energy  with ($T_1 = 0$) for a $T_2 - \Delta_2$ scan as in Figure~\ref{fig:HM1}. Note the change in colour scale showing that only negligible photon counts are recorded. (b) Measured photon counts as in (a) without the second pulse ($T_2 = 0$) at $\Delta_1=\SI{-4.9}{\milli\electronvolt}$ and  $T_1=0\ldots1$. Red dashed line shows the experimental conditions $T_1 = 0.5$ for all the results displayed in the main text.}
    \label{fig:single_pulse}
\end{figure}

To verify that the measured exciton population is the effect of the two-pulse excitation, we perform control experiments. For this, we first set $T_1 = 0$, and perform the $T_2 - \Delta_2$ scan, as described in the main text. We observe that, as shown in Figure~\ref{fig:single_pulse}(a), the measured exciton counts are insignificant. Next, we set $T_2 = 0$ (which implies there is no $\Delta_2$ scan), and simply perform a $T_1$ scan, recording the exciton counts. The results are displayed in Figure~\ref{fig:single_pulse}(b). The integrated exciton counts vary in accordance with the rise in the background, indicating yet again the absence of the SUPER effect. The red dashed line in Figure~\ref{fig:single_pulse}(b) denotes the experimental conditions for the results displayed in Figure~\ref{fig:HM1}, asserting that the measured exciton counts are only due to the two-pulse excitation. Nonetheless, for $T_1 > 0.8$, we observe a minor increase in the exciton counts, which we attribute to the substrate luminescence. Thus, for the below-band gap, SUPER excitation of the quantum dot, we strictly require two detuned pulses. 

\subsection{Effect of the first pulse intensity}
\label{app:1stPulseIntensity}

\begin{figure}[hbt]
    \centering
    \includegraphics[width = 1\linewidth]{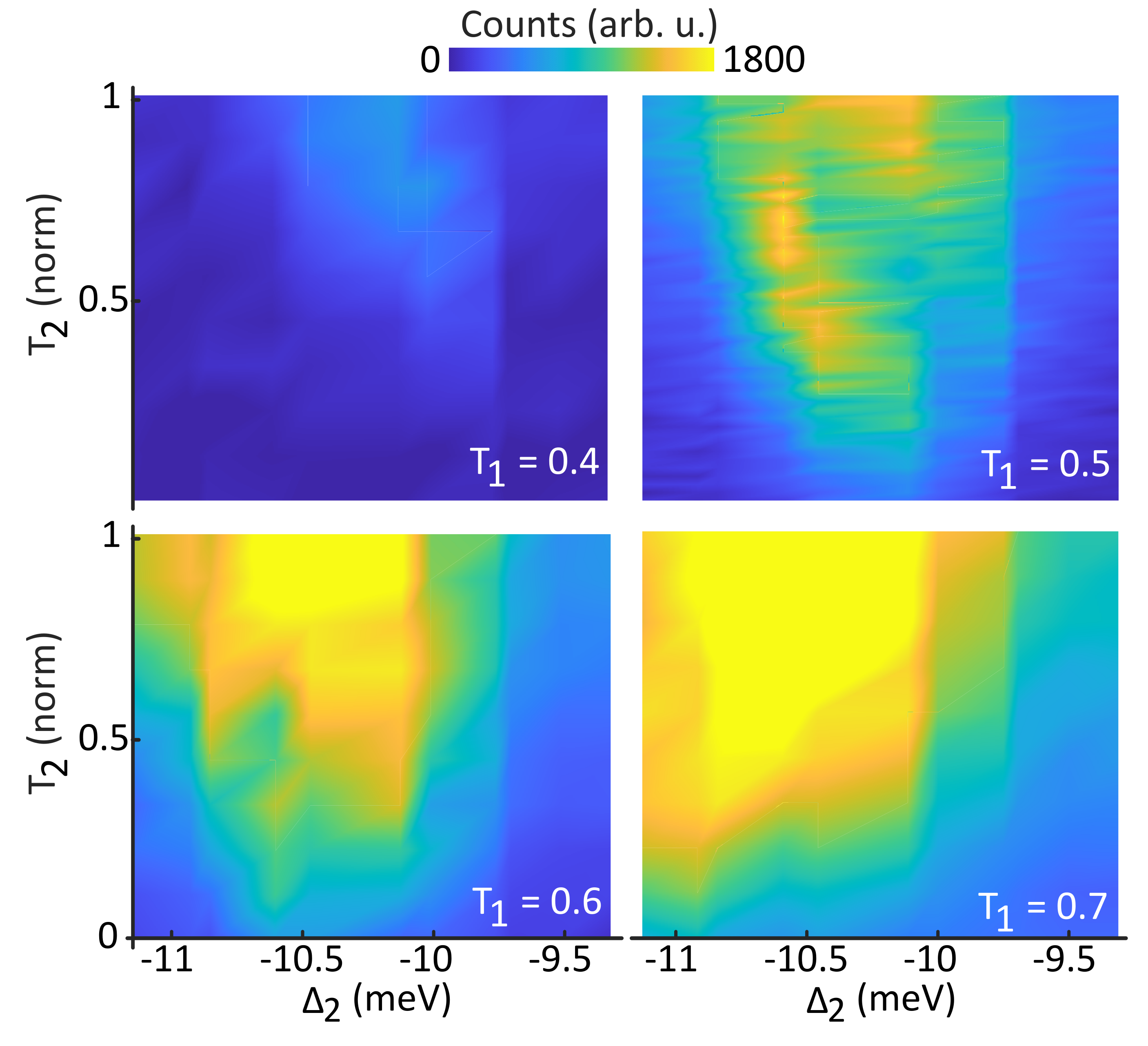}
    \caption{\textbf{Variation of exciton counts with $T_1$:} Measured population of the exciton state as function of $\Delta_2$ and $T_2$ for fixed intensities of the first pulse $T_1 = 0.4, 0.5, 0.6, 0.7$ (cf. Figure~\ref{fig:HM1}(a)).}
    \label{fig:T1_scan}
\end{figure}

Here, we investigate the influence of the first pulse intensity $T_1$ in the SUPER scheme. At first, we set $T_1=0.4\ldots0.7$, keeping $\Delta_1 = \SI{-4.9}{\milli\electronvolt}$ fixed. We then perform the $T_2 - \Delta_2$ scan as in Figure~\ref{fig:HM1}(a). Note that for the data displayed in Figure~\ref{fig:HM1}(a) we chose $T_1=0.5$. For $T_1 < 0.4$, we measure only a modest exciton photon count, indicating that the exciton state is not efficiently populated. At higher $T_1$ values, we do not observe the oscillatory trend as in Figure~\ref{fig:HM1}(c), but the increase in photon counts with $T_2$, indicates that the SUPER excitation takes place here too.

\subsection{Emission dynamics}
\label{app:LifeTime}
\begin{figure}[hbt]
    \centering
    \includegraphics[width = 1\linewidth]{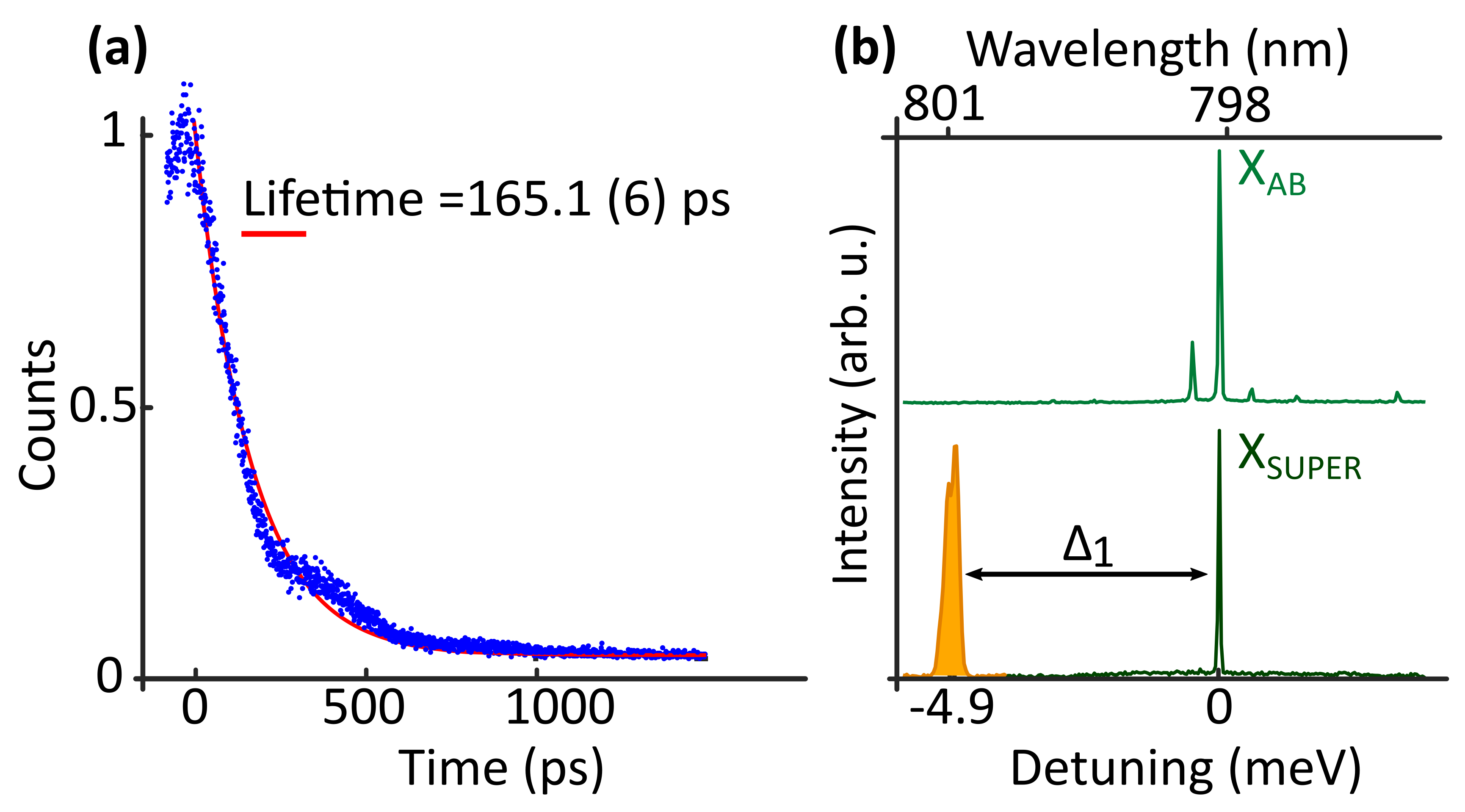}
    \caption{\textbf{The nature of the emission:} (a) Measured decay dynamics at the exciton emission energy under SUPER excitation for the conditions indicated by the red dot in Figure~\ref{fig:HM1}(a). An exponential fit to the recorded photon counts leads to a computed lifetime of \SI{165.1(6)}{\pico\second}. (b) Spectra recorded under above-band excitation contrasted to that under SUPER excitation. It is clear that above-band excitation produces spurious charge-induced emission lines (top panel, $X_{AB}$) while only a single emission line is observed under the SUPER excitation (lower panel, $X_{SUPER}$) with a \SI{3}{\nano\meter} transmission window of the bandpass filter Sec. \ref{app:background}. The orange shaded region at \SI{-4.9}{\milli\electronvolt} is the residual laser leak from the first pulse for the SUPER excitation.}
    \label{fig:decay_dynamics}
\end{figure}

To verify the nature of the emission, we investigated the decay dynamics at the exciton emission energy under the SUPER excitation, with time-correlated single photon counting as explained in the Sec. \ref{sec:Photon_Q}. The computed lifetime of \SI{165.1(6)}{\pico\second} obtained with an exponential fit to the recorded photon counts agrees with our results under the resonant excitation. We also note that, in principle, the excitation with two pulses could result in a two-photon absorption in the quantum dot or surrounding barrier, but this would result in a much slower decay. Furthermore, in Figure~\ref{fig:decay_dynamics} (b) we compare the emission spectra obtained on above-band excitation to that obtained on the SUPER excitation. It is clear that above-band spectrum contains a number of lines, attributed to charge fluctuations, which are absent in the one obtained under the SUPER excitation, asserting that the latter is a coherent excitation process.  

\end{document}